\documentclass[twocolumn,floats,showpacs,prd,amssymb,floatfix,nofootinbib,balancelastpage,superscriptaddress,amsmath]{revtex4}
\usepackage{epsfig}
\numberwithin{equation}{section}

\newcommand{\be}{\begin{equation}}
\newcommand{\ee}{\end{equation}}
\newcommand{\ds}{{\sffamily DarkSUSY}}

                              \newlength{\strikewidth}
                              \newlength{\strikelength}
\begin{document}

\title{Dark matter from late decays and the small-scale structure problems}

\author{Francesca Borzumati}
\email{borzumat@ictp.it}
\affiliation{ICTP, Strada Costiera 11, 34014 Trieste, Italy}
\affiliation{Department of Physics, National Central University, Chung-Li 32054, Taiwan}
\author{Torsten Bringmann}
\email{bringman@sissa.it}
\affiliation{SISSA/ISAS and INFN, Trieste, Via Beirut 2-4, I-34014 Trieste, Italy}
\author{Piero Ullio}
\email{ullio@sissa.it}
\affiliation{SISSA/ISAS and INFN, Trieste, Via Beirut 2-4, I-34014 Trieste, Italy}

%------------------------------------------------------------------------------

\begin{abstract}
The generation of dark matter in late decays of quasi-stable massive particles has been proposed as a viable framework to address the excess of power found in numerical N-body simulations for cold dark matter cosmologies. We identify a convenient set of variables to illustrate which requirements need to be satisfied in any generic particle physics model to address the small scale problems and to fulfill other astrophysical constraints. As a result of this model-independent analysis, we point out that meeting these requirements in a completely natural way is inherently difficult. In particular, we re-examine the role of gravitinos and Kaluza-Klein gravitons in this context and find them disfavoured as a solution to the small-scale problems in case they are DM candidates generated in the decay of thermally produced WIMPs. We propose right-handed sneutrinos and right-handed Kaluza-Klein neutrinos as alternatives. We find that they are viable dark matter candidates, but that they can contribute to a solution of the small scale problems only in case the associated Dirac neutrino mass term appears as a subdominant contribution in the neutrino mass matrix.
\end{abstract}

\keywords{cosmology}

%------------------------------------------------------------------------------
% User-supplied List of keywords.

\pacs{98.80.Cq, 98.80.Es, 95.35+d, 12.60.Jv, 14.80.Ly}

\maketitle

\section{Introduction}

Cosmological observations have given overwhelming evidence that
nonbaryonic dark matter~(DM) is the building block of structures in the
Universe~\cite{wmap}. At the same time, from a particle physics
perspective, little is known about the detailed properties of DM
particles. In the standard cosmological scenario, a cold dark
matter (CDM) term is introduced as a generic component that is 
i)~``dark" or dissipationless, as it couples to photons and baryons 
only through gravity, 
ii)~``cold'', i.e.~with negligible free-streaming effects, and
iii)~collisionless. 

This scenario can be accommodated in extensions to the standard model
(SM) of particle physics. The most attractive scheme is probably the
one in which CDM is introduced as a thermal relic component: stable
massive particles that acquire a relic density of the order of the DM
density in the Universe, provided their coupling to SM particles is 
of weak type.  The list of weakly interacting massive 
particles~(WIMPs) that 
have been proposed as DM candidates includes, among others, the 
lightest neutralino in supersymmetric models in which this is the
lightest supersymmetric particle (LSP), and the first 
Kaluza-Klein~(KK)
hypercharge gauge boson in models with universal extra dimensions 
(for reviews, see~\cite{reviews}).

In CDM cosmologies, structures form hierarchically, with small
structures collapsing first and then merging into larger and larger
bodies. This picture has proven to be remarkably successful to
describe the distribution and correlation of structures on large
scales. There is, however, some tension with observations on
small scales
in the non-linear structure formation regime which needs
to be studied with numerical N-body simulations. The focus has been,
in particular, on two issues. The first is the overabundance of
substructures in simulated halos of Milky Way size, with respect to 
the observed number of galactic
satellites~\cite{Klypin:1999uc,Moore:1999nt}. The second regards the
rise in the rotation curves of small, DM-dominated galaxies
which seems, on average, to point to profiles of DM density with
a flat inner core (see, e.g.,~\cite{FP,moorerot,BS}), as opposed to
the large concentrations and cuspy profiles found in
simulations~\cite{NFW,n04,d05}. 
The debate on whether these
discrepancies are calling for a deeper understanding of the
astrophysical and cosmological processes in connection to structure
formation, or whether they are actually pointing to a drastic change
of the CDM framework, is still open.

In both cases, the discrepancies between observations and simulations
are alleviated in schemes with a suppression of the power spectrum at
small scales. Variants to the standard cosmological model embedding
this feature, while leaving the picture at large scales unchanged,
most often involve a (mild) violation of at least one of the three
properties listed above for CDM. Proposals include: self-interacting
DM~\cite{Carlson:1992,Spergel:1999mh}, warm dark 
matter~(WDM)~\cite{Dalcanton:2000hn}, DM with a very large 
 pair
annihilation rate~\cite{Kaplinghat:2000vt}, and fuzzy 
DM~\cite{Hu:2000ke}.  In general, it is much more contrived to
construct SM extensions with particles of this kind. For example, a
sterile neutrino with a mass in the keV range, one of the most
promising candidate for WDM has recently been excluded as a thermal DM
candidate~\cite{Seljak:2006qw,Viel:2006kd}.

An alternative approach involves introducing two distinct phases in
structure formation. This is achieved by assuming that at least part
of the DM observed in the Universe today has been produced, at late
times, in the decay of a long-lived species. In such a setup two
different mechanisms can provide the needed suppression of the 
power spectrum, depending on the nature of the decaying species.

If all DM is generated in the decay of a charged species, the matter
power spectrum is essentially cut off on scales that entered the
horizon before the decay~\cite{Sigurdson:2003vy}. This is because the
charged species is tightly coupled to the photon-baryon fluid.  In
Ref.~\cite{Profumo:2004qt}, an explicit model was constructed in the
minimal supersymmetric standard model (MSSM) context. In this
model, a fraction of today's DM neutralinos is produced in the late
decay of staus, implying a scale-dependent matter power spectrum, and
in turn, a reduced power on small scales. As recently noticed,
however, long-lived charged particles may play the role of catalyzers
during big bang
nucleosynthesis~(BBN)~\cite{Pospelov:2006sc,Kaplinghat:2006qr} and a
sharp increase in the primordial Lithium abundance may be induced,
possibly in contradiction with observations.

The second mechanism is connected to the decay itself and takes
place if the produced particles have kinetic energies much larger 
than those of the corresponding thermal relic components, making 
such DM candidates warm or even hot. This mechanism, discussed first
in~\cite{Borgani:1996ag,Lin:2000qq,Hisano:2000dz}, was re-examined
recently~\cite{Cembranos:2005us,Kaplinghat:2005sy,Jedamzik:2005sx,Steffen:2006hw,Strigari:2006jf}
for DM generated in the decay of quasi-stable thermal
relic
WIMPs~\cite{Covi:1999ty,Feng:2003xh,Feng:2003uy,Ellis:2003dn,Feng:2004zu,Cerdeno:2005eu},
such as gravitinos or axinos from next-to-lightest supersymmetric
particle (NLSP), or such as KK gravitons from thermally produced 
KK states.

Here, we reconsider the metastable WIMP scenario. We start
Sect.~\ref{sec_setup} by reviewing the general idea and give a
summary of the relevant astrophysical constraints.  We then identify a
convenient set of observables and the range of values in which they
must be confined in order to solve the small-scale structure problems
of standard CDM.  The discussion is general and applies to any
particle physics framework, providing therefore a useful tool to
discriminate among different DM candidates. In
Sect.~\ref{sec_scenarios}, we analyze in particular the gravitino and
the KK graviton, that have been claimed to solve such
problems and we find them disfavoured.  As an alternative, we propose
right-handed sneutrinos and Kaluza-Klein right-handed neutrinos, and discuss in
detail their potential in this context.  In Section~\ref{sec_conc} we
present our conclusions.

%%%%%%%%%%%%%%%%%%%%%%%%%%%%%%%%%%%%%%%%%%%%%%%%%%%%%%%%%%%%%%%%%%%%%%%%%%%%

\section{Dark matter from long-lived particles}
\label{sec_setup}

We consider a setup in which today's dark matter component in the Universe, in the form of 
some particle $X$, is generated in the decay of a relic population of quasi-stable particles $Y$. Both $X$ and $Y$ are treated as dissipation-less and collision-less for what regards structure formation, and we assume that the abundance of the particles $X$ prior to the decay is negligible. Today, the DM abundance is thus given by
\be
  \label{OmegaDM}
  \Omega_X=\frac{M_X}{M_Y}\Omega_Y\,,
\ee
where $M_X$ and $M_Y$ are the respective particle masses; $\Omega_Y$ is the relic density that the species $Y$ would have acquired if it were stable. Such a two-phase DM scenario has been advocated to address both of the above-mentioned small-scale 
problems of standard CDM~\cite{Cembranos:2005us,Kaplinghat:2005sy,Jedamzik:2005sx,Strigari:2006jf}; we review here its main features.

To begin with, the very steep central cusps found in CDM simulations of DM halo profiles tend to be smoothed 
out in this setup: the particle $X$ picks up an increased velocity dispersion in the 
decay, and the DM phase-space density, $Q\propto \rho/\langle v^2\rangle^{3/2}$, is significantly 
reduced as compared to the standard CDM case. 
Following \cite{Kaplinghat:2005sy,Strigari:2006jf}, we consider an average measure of the primordial DM phase-space density that is obtained by a full integration over the underlying phase-space distribution function:
\be
  Q_p\equiv\frac{\bar \rho}{\bar\sigma^3}=10^{-24}\alpha\,\left(\frac{M_X}{p_\mathrm{cm}a_d}\right)^3\,,
\ee
where $\alpha=1.0$ (0.8) for decays in the radiation (matter) dominated era,  $p_\mathrm{cm}$ denotes the center of mass momentum of the daughter particle $X$ after the decay 
and $a_d$ is the cosmic scalefactor at decay. The above fine-grained value of $Q_p$ has the 
property of remaining constant for dissipation-less matter, while its coarse-grained version can 
only decrease. The assumption that the evolution of a galaxy is mostly dominated by its dark 
(i.e. dissipation-less) component, would thus imply the lower bound $Q_p\gtrsim Q_0$, with
$Q_0$ the largest coarse-grained value observed in galaxies, about 
$Q_0\equiv 10^{-4}\left(M_\odot/\mathrm{pc}^3\right)\left(\mathrm{km}/\mathrm{s}\right)^3$~\cite{Dalcanton:2000hn}. On the other hand,
the effect of baryons, most probably, cannot be neglected in real galaxies, and primordial 
phase-space densities as small as $Q_p\sim10^{-2}Q_0$ have been considered \cite{Strigari:2006jf},
corresponding to the \emph{average} central phase-space density that is observed in low-mass spirals. Independently of whether one actually can argue for the existence of a strict lower bound, $Q_p$ 
has to  satisfy
\be
  \label{Qbound}
  Q_p\lesssim Q_0
\ee
in order to reduce the cusps in simulated halos and match the shallower profiles observed 
for low-mass objects~\cite{Dalcanton:2000hn,Strigari:2006jf}. Note that the above analyses are based on (semi-) analytical considerations. While numerical simulations with decaying dark matter particles do not exist at the moment, it would be interesting to study in this way the actual evolution of $Q_p$ and compare the results with the naive expectations.

The second main feature of the two-phase DM scenario considered here is the introduction of a net free-streaming 
effect which may in general be much larger then in the case of thermally generated WIMPs. 
In Ref.~\cite{Kaplinghat:2005sy} the Boltzmann equation for the decaying system is solved 
and, correspondingly, a damped matter power spectrum is derived. For our purposes, however, it will be sufficient 
to refer to the free-streaming length
\be
  \label{def_fs}
  \lambda_\mathrm{FS}\equiv\int_{\tau}^{t_0}\frac{v_X(a)}{a}dt\,,
\ee
where ${\tau=t(a_d)}$ is the lifetime of the decaying particle $Y$, and $v_X=(p_\mathrm{cm}/M_X)(a_d/a)$ the 
velocity of the daughter particle $X$. $\lambda_\mathrm{FS}$ gives, approximately, the scale below 
which primordial perturbations are erased. In current cosmological data there is no direct
evidence for a departure of the matter power spectrum from the standard $\Lambda$CDM 
form; recent limits on WDM setups, derived using the latest Lyman-$\alpha$ 
Forest data from the Sloan Digital Sky Survey~\cite{McDonald:2004eu}, are given in 
terms of a lower bound on the mass of a sterile neutrino of about 
10~keV~\cite{Seljak:2006qw,Viel:2006kd}. With the above definition of the free-streaming scale, (\ref{def_fs}), this corresponds to an upper bound of roughly $\lambda_\mathrm{FS}\lesssim0.5\,\mathrm{Mpc}$~\cite{Strigari:2006jf}. On the other hand,
 WDM models with a 
free-streaming scale very close to this, about 
\be
  \label{lambdabound}
  \lambda_\mathrm{FS} \gtrsim 0.3\,\mathrm{Mpc}\,,
\ee
are needed in order to produce Milky Way-size galaxies with satellite populations in fair agreement with the number 
of satellites observed in our own Galaxy and in Andromeda, and thus resolve the present disagreements \cite{Colin:2000dn,Zentner:2003yd,Cembranos:2005us}.

\begin{figure}[t!]
     \includegraphics[width=\columnwidth]{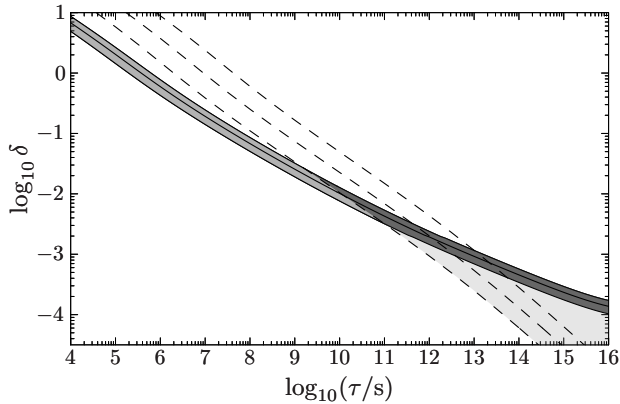}
     \caption{From bottom to top, the solid (dashed) lines correspond to $\lambda_\mathrm{FS}/\mathrm{Mpc}=0.3,0.4,0.5$ ($Q_p/Q_0=1,0.1,0.01$). In the dark shaded region both small scale structure problems could be resolved, while in the lighter shaded areas this is true for only one of them, respectively. The upper-right part is excluded from Lyman-$\alpha$ forest measurements of the power spectrum, while in the lower-left part of the plot, the mechanism of dark matter generation through 
the decay of a meta-stable species does not leave an observable imprint  in the sky.      
}
     \label{fig_Qlambda}
\end{figure} 

Having identified $Q_p$ and $\lambda_\mathrm{FS}$ as the relevant observables to address the small scale structure problems of standard CDM cosmology, we can now conveniently rephrase our 
discussion in terms of quantities which are related to the particle physics setup only: the lifetime $\tau$ of the 
decaying particle $Y$ and the mother-daughter mass splitting $\delta\equiv(M_Y-M_X)/M_X$. This is done in Fig.~\ref{fig_Qlambda}, where we map a few values of $Q_p$ and $ \lambda_\mathrm{FS}$ into the $\tau$ - $\delta$ plane. 
The regime at large lifetimes and sizable
mass splittings (upper right region in the plot) is excluded because it corresponds to too large 
free-streaming lengths, while within the shaded areas at least one of the two conditions (\ref{Qbound}) 
and (\ref{lambdabound}) is satisfied.\footnote{Note that both conditions, (\ref{Qbound}) and (\ref{lambdabound}), are rather conservative in their claim to solve the respective small scale structure problem; tightening them decreases the shaded areas of Fig.~\ref{fig_Qlambda} even further. The recent analysis \cite{Viel:2007mv}, for example, finds a considerably tighter Lyman-$\alpha$ constraint on $\lambda_\mathrm{FS}$ than what was found in \cite{Seljak:2006qw,Viel:2006kd}.} 
In the lower left part of the plot, on the other hand, there is 
no impact on the small scale structure problems at all.
Note that for $\tau\ll t_\mathrm{eq}\approx 2.0\times10^{12}\,\mathrm{s}$ the contour lines are essentially parallel,~i.e. $\lambda_\mathrm{FS}=\lambda_\mathrm{FS}(Q_p)$, while for larger lifetimes there is an additional dependence $\lambda_\mathrm{FS}=\lambda_\mathrm{FS}(Q_p,\tau)$; this effect has been stressed already in Ref. \cite{Strigari:2006jf} to point out that even very late decays, with $\tau\gtrsim t_\mathrm{eq}$, could provide a solution to the small-scale problems.

\begin{figure}[t!]
      \includegraphics[width=\columnwidth]{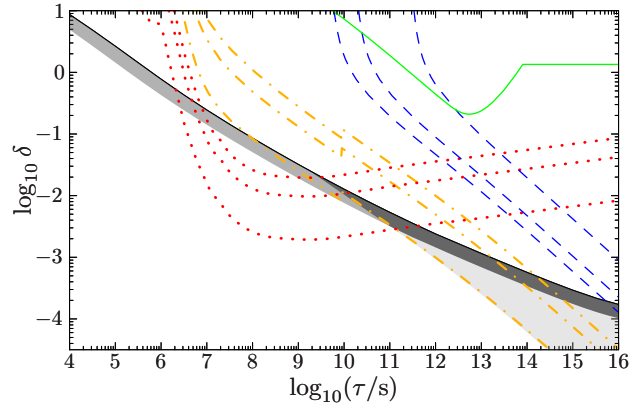}
     \caption{This plot shows the tightest available constraints on a scenario with DM from late decays. The yellow dashed-dotted (red dotted) lines give the limits from CMB (BBN), from bottom to top, when a fraction $f_\gamma=0.05,0.01,0.005\,$ of the total energy in light decay products is released into electromagnetically active species. The contribution to the ISW excludes the area above the green solid line. Finally, the blue dashed lines show, from bottom to top, the Super-Kamionkande limit for masses $M_Y=1,0.5,0.1$ TeV. The shaded regions are the same as in Fig.~\ref{fig_Qlambda}. Not included in the figure are constraints from hadronic decay products, which become important for $\tau\lesssim10^8\,$s; see text for further details.
}
     \label{fig_constraints}
\end{figure}

So far, we have focussed on the effects of the daughter DM particle $X$; in the decay of the mother particle $Y$, however,
potentially observable effects may also be induced by the fraction of energy that is released into 
light species. In particular, a late injection of relativistic energy could -- regardless of the emitted species -- be spotted in the CMB or in large scale structure (LSS) surveys. For decays before recombination, the main constraint of this type arises from CMB distortions due to the early integrated Sachs-Wolfe (ISW) effect  \cite{Zentner:2001zr}; the combined analysis of CMB and LSS data leads to an upper bound on the excess relativistic energy density at recombination that can be expressed in terms of the effective number of light neutrino species as $\Delta N^\mathrm{eff}_\nu\lesssim1.6$ \cite{Hannestad:2006mi}. For decays after the time of recombination, it is again the (late) ISW effect that is most important; it puts a bound of $M_\nu\lesssim10\,\mathrm{eV}$ on a decaying massive neutrino for $10^{13}\lesssim\tau\lesssim10^{16}$ \cite{Kaplinghat:1999xy}, which translates into $\delta\lesssim1.3$ in our scenario.\footnote{Note that in this analysis the recent WMAP data are not taken into account; the projection  of Ref.~\cite{Hannestad:1999xy} was that  they would allow to tighten the bound to $M_\nu\lesssim0.3\,\mathrm{eV}$ ($M_\nu\lesssim1.0\,\mathrm{eV}$) at $\tau\sim10^{14}\,\mathrm{s}$ ($\tau\sim10^{16}\,\mathrm{s}$), corresponding to $\delta\lesssim0.03$ ($\delta\lesssim0.1$).}
These bounds, shown as a green solid line in Fig.~\ref{fig_constraints}, cover a portion of the parameter space that is already excluded by the Lyman-$\alpha$ forest limit. A release of relativistic energy at very late times, finally, is in principle also constrained by recent supernovae data, but the corresponding bounds are even weaker than those from the ISW \cite{Zentner:2001zr}.

Turning to explicit models for the emitted light species, and excluding for the moment the introduction of new light exotic particles (such as hypothetical light scalars having escaped detection at accelerators), there are basically three possibilities:
the decay produces -- on top of the DM particle $X$ -- electromagnetic radiation, hadronic species or neutrinos. For late decays, in particular, the greatest concern usually is whether a sizable fraction 
$f_\gamma$ of the total energy in light decay products is carried away by electromagnetically active species; this 
would potentially lead to spectral distortions of the CMB \cite{Hu:1993gc} or to a spallation of the 
light elements produced during BBN (here, the tightest constraint derives from the $^6$Li 
abundance \cite{Jedamzik:2006xz}). For the CMB, the corresponding bounds in the $\tau$ - $\delta$ plane are shown in Fig.~\ref{fig_constraints} as yellow dash-dotted lines; they are derived as an update of the analysis of Ref.~\cite{Hu:1993gc} by taking into account the most recent limits on deviations of the CMB from a thermal spectrum, $|\mu_0|<9\times10^{-5}$ and $|y|<1.2\times10^{-5}$ \cite{Hagiwara:2002fs}. 
Here, we do not take into account the more refined (and slightly less constraining) CMB limits for $\tau\lesssim10^9\,$s from the recent analysis of Ref.~\cite{Lamon:2005jc}, since in this regime the BBN bounds (shown as red dotted lines) are much tighter.
As it can be seen from the figure, the entire region in the parameter space in which both small scale structure problems are solved \emph{simultaneously} 
is ruled out for $f_\gamma\gtrsim5\%$; in fact, very late decays ($\tau\gtrsim10^{14}\,$s) may be ruled out even for $f_\gamma\lesssim1\%$.
(We note in passing that for very small mass differences and decays close to today, too late to influence structure formation in the way we are interested in here, distortions of the diffuse extragalactic photon background may become more constraining than the CMB constraints shown in Fig.~2 \cite{Cembranos:2006gt,Cembranos:2007fj,Yuksel:2007dr}).
The emission of a hadronic component, secondly, would also have
observable effects, mainly in connection with the light element abundances (see, 
e.g.,~\cite{Jedamzik:2006xz} for a recent analysis). For $\tau\gtrsim10^8\,$s, the corresponding constraints turn out to be essentially the same as for the electromagnetic case (though slightly weaker), with $f_\gamma$ exchanged by the corresponding hadronic fraction $f_h$ . At earlier times, $10^{-1}\,\mathrm{s}\lesssim\tau\lesssim10^8\,\mathrm{s}$, the bounds on hadronic decays are usually considerably stronger; however, since they are also slightly more model dependent and in any case lie outside the dark shaded region, they are of limited interest to our analysis and therefore not shown here.
Finally, loose bounds are obtained in the case of neutrinos as light species in the decay;
the most stringent constraint here is usually given by the upper bound on the (electron) neutrino background flux as measured by the Super-Kamionkande experiment, $\Phi_{\nu_e}<1.2\,\mathrm{cm}^{-2}\mathrm{s}^{-1}$ for $E_{\nu_e}>19.3\,$ MeV \cite{Malek:2002ns}. This bound (blue dashed curves in the figure), however, is almost
always less restrictive than the Lyman-$\alpha$ forest limit. Other constraints, such as the one from a conversion of the emitted neutrinos into electromagnetic radiation through scattering on the 
relic neutrino background, are considerably weaker and do not approach the region of the parameter space that is of interest in our context.

Having in mind a particle physics modelling of the transition between the two dark matter phases, we can now take our discussion one step further. We have shown above that a solution to the small scale problems requires rather small mass splittings, $\delta\lesssim0.02$, and that $f_\gamma$ (as well as $f_h$) needs to be at the per cent level or lower. The second 
condition points to a model with the two-body final state $X$-neutrino as dominant decay mode. 
In fact, if this mode is allowed, by conservation of charge and lepton number, 
the emission of electromagnetic radiation is forbidden at the same level in perturbation theory; while in general possible through a three-body decay, the corresponding branching 
ratio will naturally be below $1\%$.  For a two-body decay with a spin-1/2
particle in the final state (the neutrino), we can distinguish between two phenomenologically 
interesting cases. In the first case, both $X$ and $Y$ have spin less or equal to 1 and the decay rate for small 
$\delta$ scales as:
\be
  \label{Gamma1}
  \Gamma=\frac{\left|g_\mathrm{eff}\right|^2}{8\pi} M_Y \delta^2\,.
\ee
Here, $g_\mathrm{eff}$ is an effective coupling that depends on the particular particle physics 
model and usually contains higher order corrections in $\delta$. The phase 
space integration gives a term proportional to the mass splitting; an additional factor of $\delta$ arises due to 
the light spin-1/2 particle in the final state. If, on the other hand, either $X$ or $Y$ have 
spin $3/2$ or $2$ -- prototype examples being the gravitino and the Kaluza-Klein graviton, 
respectively -- the decay rate for small mass splittings takes the form
\be
  \label{Gamma2}
  \Gamma=\frac{\left|\tilde g_\mathrm{eff}\right|^2}{3\pi M_\mathrm{Pl}^2} M_Y^3 \delta^4\,,
\ee
where $M_\mathrm{Pl}\equiv(8\pi G)^{-1/2}$ is the reduced Planck mass and one generically expects $\tilde g_\mathrm{eff}\sim\mathcal{O}(1)$. As we will show,
in both cases $M_Y$ is essentially fixed by the requirement to obtain the right DM relic density (\ref{OmegaDM})
as measured by WMAP \cite{wmap}, so the relevant free parameters are just 
$\delta$ and $g_\mathrm{eff}$ ($\tilde g_\mathrm{eff}$).

\begin{figure}[t]
       \includegraphics[width=\columnwidth]{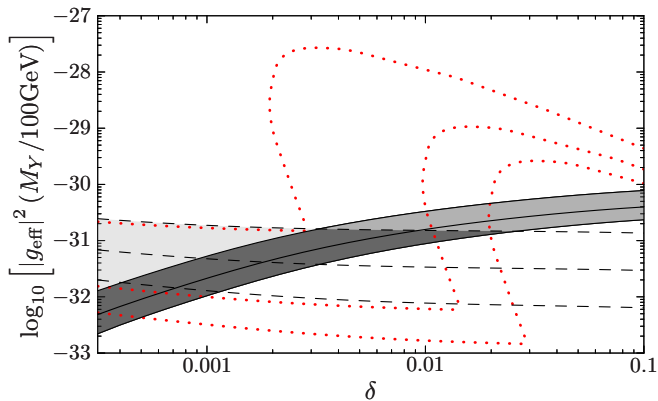}
     \caption{The required coupling strength as a function of the mass splitting, for the case that both $X$ and $Y$ have spin 0, 1/2 or 1. From top to bottom, the solid (dashed) lines correspond to $\lambda_\mathrm{FS}/\mathrm{Mpc}=0.3,0.4,0.5$ ($Q/Q_0=1,0.1,0.01$). The lower part of this plot is thus excluded and the dark shaded area shows the region where both small scale structure problems can be resolved. Also included is the combined bound from BBN and CMB, shown as a red dotted line for $f_\gamma=0.05,0.01,0.005$. 
}
     \label{fig_particle}
\end{figure} 

\begin{figure}[t]
       \includegraphics[width=\columnwidth]{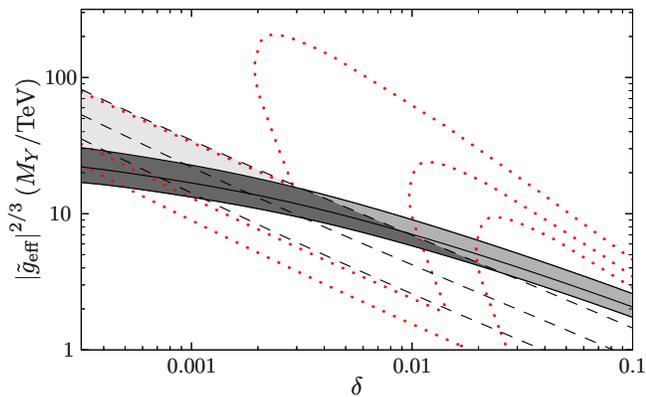}
     \caption{Same as Fig.~\ref{fig_particle}, now for the case that either $X$ or $Y$ is a spin 3/2 or 2 particle. See text for further details. 
}
     \label{fig_particle_b}
\end{figure} 

In Fig.~\ref{fig_particle} and \ref{fig_particle_b} we show the favored regimes and the constraints on the two-phase DM
scenario, rephrased in terms of the relevant particle-physics parameters; these plots allow for a quick and easy
check on whether a particular particle-physics model, with given couplings and mass spectrum, meets the requirements for 
a solution to the small-scale problems. At the same time, they illustrate that a certain amount  of 
fine-tuning in $\left|g_\mathrm{eff}\right|^2$ or $\left|\tilde g_\mathrm{eff}\right|^2$ is necessary in 
order for the scenario to work; we will elaborate further on this point in the next Section, where we introduce and discuss in more detail
some explicit models.

\section{Viable particle physics scenarios}
\label{sec_scenarios}

As a starting point to classify viable frameworks, we consider the various mechanisms that may guarantee a long lifetime for the state $Y$. In Ref.~\cite{Lin:2000qq}, e.g., to start with,
cosmic strings are assumed to play the role of the metastable state $Y$: in general, this scenario
requires tunings both on the abundance of the decaying species and on the decay products.
Another possibility is explored in Ref.~\cite{Profumo:2004qt}, where the decay rate
is suppressed because the allowed phase space gets sharply reduced,
such as in the case of a multi-body final state in the limit of small mass splittings $\delta$; this 
mechanism becomes viable only in case of an electrically charged particle in the initial state,
and hence does not fit into the scenario we have described in the previous Section.

Here, we will instead consider a framework where quasi-stable states arise due to strongly suppressed couplings, i.e.~when one of the particles involved in the decay is 
super-weakly interacting (a "superWIMP", as it is sometimes dubbed in the literature).
At first sight,  it is thus not important whether the superWIMP appears in the initial or  in the
final state. On the other hand, supposing that the particle $Y$ in the initial state is a WIMP
(and hence that $X$  in the final-state is a superWIMP)  allows to invoke thermal production 
as a natural mechanism for the generation of the dark matter component of the Universe.
The reverse is less appealing for two reasons: First, it requires some non-thermal 
mechanisms to generate initial-state superWIMPs. At the same time, in explicit models, 
WIMP DM appears as the lightest species in a tower of extra particles sharing a common 
quantum number, usually with moderate mass splittings among these states;
in general, it seems unnatural to add a superWIMP as next-to-lightest particle
in this construction and to suppose that the thermal relic density of the WIMP is
completely neglible. In the following, we will therefore focus on thermally generated WIMPs as decaying species, and 
refer to the standard cosmological setup to estimate the relic densities at the time of decay. 
In principle, there could be non-standard (effective) contributions to the total energy density of the universe at the decoupling epoch (anticipating the freeze-out process) that enhance the relic abundance
(see, e.g. ~\cite{Griest:1990kh,Salati:2002md,Profumo:2003hq}), or entropy releases after 
decoupling  which would dilute the relic population~\cite{Griest:1990kh}. In both cases, however, 
we would loose predictability and, to some extent,  simply reduce to a scenario in which 
the abundance of the decaying species is  fine-tuned such as to match the effect we wish to account for;
we prefer to avoid such a setup.

To enforce a two body decay of a WIMP into a superWIMP and a neutrino, we exploit lepton 
number conservation and impose that either the WIMP or the superWIMP carries lepton number.
We start within a supersymmetric setup, where we encounter two possibilities: gravitino dark matter produced
in sneutrino decays~\cite{Feng:2004zu,Buchmuller:2006nx,Kanzaki:2006hm}, or right-handed 
sneutrino dark matter produced in neutralino decays~\cite{Asaka:2005cn,Gopalakrishna:2006kr}. Later, in Subsection \ref{UED}, we will discuss the corresponding cases in scenarios with universal extra dimensions.

\subsection{Gravitino dark matter from left-handed sneutrino decays}

In the minimal supersymmetric extension to the Standard Model (MSSM), 
left-handed sneutrinos acquire their masses through a soft SUSY-breaking term
and a D-term, i.e. $M_{\tilde\nu}^2 = M_L^2 + D_\nu$.  Analogously, a mass term for
left-handed charged sleptons is generated: $M_{\tilde{l}_L}^2 = M_L^2+  m_l^2 + D_{l}$,
where $m_l$ is the mass of the corresponding lepton. The D-term contributions are, respectively, 
$D_\nu = 1/2 \, m_Z^2 \cos 2\beta$ and $D_{l} = (-1/2+\sin^2\theta_W) \, m_Z^2 \cos 2\beta$,
where the angle $\beta$ is  defined by the ratio of the vacuum expectation values of the 
two MSSM Higgs doublets as $\tan\beta \equiv \langle H_2^0 \rangle / \langle H_1^0 \rangle$.
Since $\tan\beta >1$,  $D_\nu$ is a negative correction, while $D_l$ is positive; one thus finds 
that sneutrinos are lighter than the corresponding left-handed charged sleptons. 
If, furthermore, the right-handed charged slepton soft terms are larger than their left-handed counterparts (for simplicity, we will 
always assume in the following that they are, in fact, much larger; right-handed charged sleptons then effectively decouple from the theory) 
the lightest sneutrino is the lightest slepton, and possibly the lightest particle among SUSY 
counterparts of SM fields. We focus on this case, and estimate the thermal relic abundance 
of the quasi-stable sneutrinos after freeze-out. Then, as a second step, disjoint from thermal 
decoupling since we work under the hypothesis of quasi-stability, we consider sneutrino decays 
into the lightest (and stable) supersymmetric particle (LSP), which we assume to be the gravitino:
\be
\tilde{\nu} \rightarrow \tilde{G} + \nu\,.
\ee
The decay width for this process is \cite{Feng:2004zu}:
\be
  \Gamma=\frac{1}{96\pi M_\mathrm{Pl}^2} \frac{M_{\tilde{\nu}}^5}{M_{\tilde{G}}^2}
  \left[ 1 - \left(\frac{M_{\tilde{G}}}{M_{\tilde{\nu}}}\right)^2 \right]^4\,,
\ee
i.e. it takes the same form as in Eq.~(\ref{Gamma2}), with
\be
  \left|\tilde g_\mathrm{eff}\right|^2 = \frac{(1+\delta/2)^4}{(1+\delta)^6}\,.
\ee
As anticipated, the decay rate depends only on $M_{\tilde\nu}$ and $\delta$; however, since 
we are focussing on the case of thermally produced sneutrinos, these are, in fact, not 
free parameters  but both correlated to the amount of dark matter observed in the
Universe today. 

\begin{figure}[t!]
   \includegraphics[width=\columnwidth]{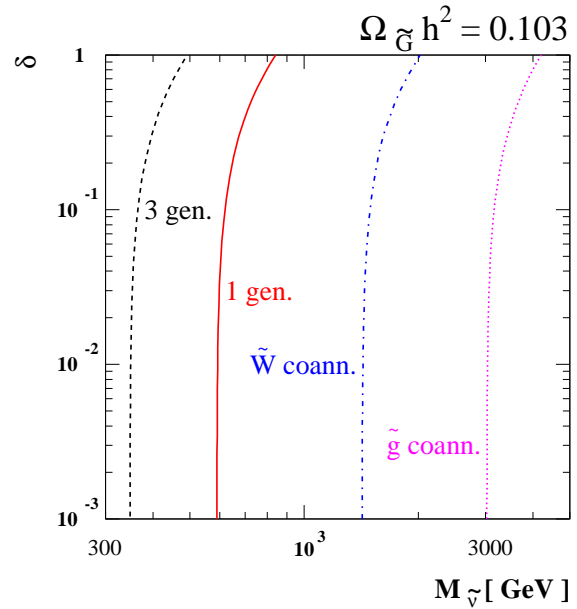}
     \caption{Correlation between sneutrino mass $M_{\tilde \nu}$ and sneutrino-gravitino mass splitting $\delta$
     for models matching the dark matter abundance measured by WMAP. Predictions within the MSSM
     span from the case of 3 families of mass-degenerate left-handed sneutrinos (dashed curve) to
     the case of coannihilations of 1 light sneutrino with gluinos (dotted curve). Also shown are
     the cases of 1 light sneutrino coannihilating with Winos and that of 1 light sneutrino family without
     coannihilations (other than with the light charged sleptons).}
     \label{fig:snue}
\end{figure} 

\begin{figure}[t!]
   \includegraphics[width=\columnwidth]{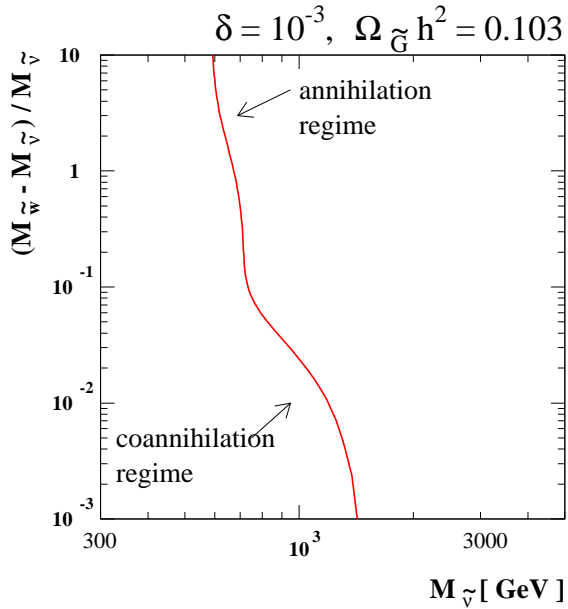}
     \caption{Correlation between sneutrino mass and sneutrino-Winos mass splitting for models 
     matching the dark matter abundance measured by WMAP, in case of small sneutrino-gravitino 
     mass splitting and of 1 generation of light left-handed sleptons}
     \label{fig:snue2}
\end{figure} 

In the following, rather than discussing this situation for the fully general MSSM, we will focus on a few specific cases that serve to illustrate the main trends to be expected in the general case. The minimal setup from that perspective is the one in which 
we assume that only the parameter $M_L$ for one single lepton family is light, while all other 
SUSY parameters are at a heavier scale (in actual calculations we choose this scale to be 10~TeV). 
The sneutrino relic density is then set by the sneutrino pair annihilation rate plus the coannihilation 
with the left-handed charged slepton; it scales approximately with the inverse of the effective 
thermally-averaged annihilation cross section: 
\be
  \Omega_{\rm thermal} \propto \frac{1}{ \langle \sigma_{\rm{eff}} v \rangle}\,,
\ee
where $\langle \sigma_{\rm{eff}} v \rangle$ includes all annihilation and coannihilation 
processes, properly waited. For an accurate estimate of the relic abundances one needs to solve
a system of coupled Boltzmann equations; we follow the approach of Ref.~\cite{Edsjo:1997bg} 
and perform the necessary numerical calculations with the  \ds\  package~\cite{Gondolo:2004sc}.
The result is shown in Fig.~\ref{fig:snue}, where we plot as a solid line the values for $M_{\tilde\nu}$ and $\delta$ that correspond to the best 
fit value of $\Omega_{DM}$ as obtained from the WMAP data~\cite{wmap}; the scaling between 
gravitino and sneutrino density is simply given by Eq.~(\ref{OmegaDM}).
Another possibility is that all three families of left-handed sleptons are light.
In Fig.~\ref{fig:snue}, we have included this situation as a dashed line, assuming the same $M_L$ for the three generations; as expected, the resulting lightest sneutrino relic abundance becomes a factor of about 3 larger than in the previous case (it is not exactly 3 times larger, since now there are  
more coannihilation processes involved), and this has to be compensated for by a larger effective annihilation rate, i.e. by a lighter sneutrino mass. 

Coannihilations with species that have much larger annihilation rates than sneutrinos, on the other hand, will
tend to shift the mass-range of interest for a  thermal DM 
production to more massive sneutrinos. As an example, we consider a framework in which
the Wino mass parameter $M_2$ is the lightest gaugino soft SUSY-breaking term, so that the
Wino-like neutralino and chargino become the lightest fermionic SUSY particles: such a scenario is
predicted in anomaly-mediated SUSY breaking schemes~\cite{Randall:1998uk,Giudice:1998xp}, 
or could emerge, e.g., in supergravity frameworks with non-universal GUT gaugino 
masses~\cite{Baer:2000gf}. Winos have very large pair annihilation rates into gauge bosons; 
in case they are the LSP, and excluding degeneracies in mass with other SUSY particles, 
their thermal relic density matches the measured DM density for a mass of about 2.2~TeV  
(see, e.g., \cite{Masiero:2004ft}). The dash-dotted line in Fig.~\ref{fig:snue} corresponds to the
case of a 0.1~\% mass splitting between Winos and the sneutrino LSP (assuming one light left-handed 
slepton generation). Wino coannihilation 
effects keep the sneutrinos in thermal equilibrium for a longer time as compared to the previous cases; the relic abundance is 
accordingly reduced, which needs to be compensated for by shifting the sneutrino mass scale 
up to values larger than about 1.4 TeV. The dependence of this result on the Wino-sneutrino mass 
splitting is shown in Fig.~\ref{fig:snue2}: we can see that the sneutrino relic density
(and hence the inferred sneutrino mass) changes significantly in the coannihilation regime -- which can be understood from the fact that the mass splitting enters exponentially in the coannihilation 
contributions to the thermally averaged effective annihilation rate $\langle \sigma_{\rm{eff}} v \rangle$. When the mass splitting
is larger than about 10\%, coannihilation effects become negligible; there is still some 
dependence of the result on the Wino mass since neutralinos and charginos 
mediate t- and u-channel annihilations of sleptons into leptonic final states and therefore 
enhance the slepton annihilation rate when they are light. 

We have just demonstrated that coannihilation effects may play a major role in determining the mass
scale for thermal relic sneutrinos. We will now, by taking a closer look at this point, infer an upper bound on the sneutrino mass for any 
given sneutrino--gravitino mass splitting (within the MSSM). Let ${\tilde c}$ be a coannihilating particle 
with a pair annihilation cross-section much larger than that of 
slepton pairs: $(\sigma v)_{{\tilde c} {\tilde c}} \gg \sum (\sigma v)_{{\tilde l} {\tilde l}^\prime}$ 
(for simplicity, we assume S-wave annihilations and take the limit of  zero relative 
velocity for initial state pairs). In the extreme case of an exact mass degeneracy among all coannihilating 
particles, the sneutrino relic abundance scales down to (see, e.g., \cite{Profumo:2006bx}):
\be
  \left(\Omega_{\tilde \nu} h^2\right)_{{\rm with}\; {\tilde c}} 
  \sim \frac{\sum (\sigma v)_{{\tilde l} {\tilde l}^\prime}}{(\sigma v)_{{\tilde c} {\tilde c}}} 
  \left( \frac{\sum g_{\tilde l} + g_{\tilde c}}{g_{\tilde c}}\right)^2 
  \left(\Omega_{\tilde \nu} h^2\right)_{{\rm without} \;{\tilde c}}\,,
\ee
where $g_i$ denotes the number of degrees of freedom for the particle $i$ (The $g_i$  in this expression account for the mismatch between the states 
maintaining thermal equilibrium and the total number of states contributing to the relic abundance 
after decoupling). The largest effect is thus obtained for the coannihilating particle with the largest
pair annihilation rate per degree of freedom, which in the MSSM is given by the gluino. For illustration, we plot in Fig.~\ref{fig:snue} as a dotted line the case of a 0.1\% gluino--sneutrino mass splitting (again for only one
"light" generation of left-handed sleptons).  For small $\delta$, we then find that $M_{\tilde \nu}$ is about 
3~TeV. In fact, this can be interpreted as a strict upper bound on the mass of a thermally produced sneutrino: advocating a further increase in the
effective thermal annihilation cross-section, such as for an extreme S-channel resonance,
is hardly plausible since even in less minimal frameworks it seems hard to introduce
stronger interacting states. Note also that this upper bound is even more sensitive to the 
mass splitting between the sneutrino and the coannihilating state than what we have shown in Fig.~\ref{fig:snue2} 
for the case of Winos; the analogous plot for a gluino would have a much sharper transition out of
the coannihilation regime, since the coannihilation effect is stronger and gluinos do not 
enter in any way into slepton annihilation processes.

\begin{figure}[t]
       \includegraphics[width=\columnwidth]{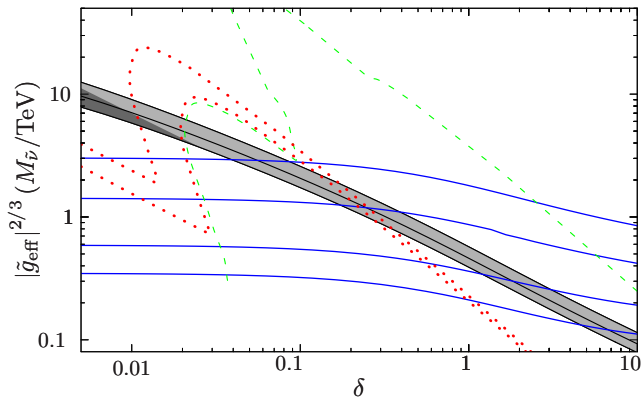}
     \caption{Projection of the viable $M_{\tilde\nu}$--$\delta$ configurations
     into the plane that we introduced in Fig.~\protect{\ref{fig_particle_b}} and here extend
     to larger values of $\delta$. The four solid lines, 
     almost horizontal in the small $\delta$ limit, correspond to the cases considered 
     in Fig.~\protect{\ref{fig:snue}}; the top curve corresponds to the configuration with the highest possible sneutrino mass (where the sneutrino relic density is mainly determined by coannihilations with gluinos). Favored and excluded regions are shown with the same coding as in 
     Figs.~\protect{\ref{fig_particle}} and~\protect{\ref{fig_particle_b}}. Finally, the BBN limit
     for a hadronic fraction  $f_h=0.005$ has been added as a dashed line.}
     \label{fig_gravitino}
\end{figure} 

Now that we have obtained an understanding of the viable $M_{\tilde\nu}$--$\delta$ configurations in our setup, we are in a position to refer to our general discussion in Section \ref{sec_setup} in order to assess the potential impact of a gravitino SuperWIMP DM candidate on the small scale problems of standard CDM cosmology. To this end, we present in Fig.~\ref{fig_gravitino} the gravitino configurations discussed above in the context of a rescaled version of Fig.~\ref{fig_particle_b}, which we have argued before to be most convenient for this type of assessments.
The four solid lines, nearly horizontal in the small $\delta$ limit, 
correspond to the four cases considered in Fig.~\ref{fig:snue} (smallest $M_{\tilde\nu}$
at the bottom, largest at the top) and span the full range of possibilities within the MSSM. 
As we can see, there is actually no model that falls into the dark shaded region, where both small-scale 
problems could be resolved -- to do so, higher sneutrinos masses would be required, which, however, is inconsistent with a thermal production scheme. Note, furthermore, that a large portion of configurations is located below the 
light shaded region; these configurations are excluded as DM scenarios since they are associated to a free-streaming scale incompatible with Lyman-$\alpha$ measurements of the matter power spectrum. Also included in Fig.~\ref{fig_gravitino} are the constraints corresponding to a 1\% or 0.5\% electromagnetic branching ratio (dotted curves), as well as a 0.5\% hadronic branching ratio (dashed curve), respectively. These values are at the level 
of what is generically expected from subdominant channels in sneutrino to gravitino decays \cite{Feng:2004zu}.
Most models within the light shaded region, that would potentially solve
the problem of the overabundance of substructures in Galactic-size halos, are thus excluded 
by the constraints from BBN. This is the general trend; in a 
model by model comparison against BBN constraints, 
see~\cite{Feng:2004zu,Buchmuller:2006nx,Kanzaki:2006hm}, it may be possible to find some configurations in the light shaded region that are not excluded since they are fine-tuned to prevent the emission of hadrons even in sub-leading processes; such a detailed analysis, however, is beyond the scope of the present work.

\subsection{Right-handed sneutrino dark matter from neutralino decays}

Having learned that gravitino DM cannot contribute to a solution of both small scale problems (in fact, it can hardly account for even one of them), we now try to reverse the picture and turn to an example where the final state superWIMP is
the particle carrying the lepton number, while the initial state is a lepton-flavor neutral WIMP. 
In SUSY frameworks, this is possible if we introduce a right-handed sneutrino as the LSP,
and impose that the lightest neutralino is the next-to-lightest SUSY particle (NLSP).

The definition of an extension to the MSSM including right-handed neutrino superfields 
$\hat{N}_{R}$ is straightforward. The minimal setup involves just one extra term in the 
superpotential (see \cite{Asaka:2005cn,Gopalakrishna:2006kr}):
\be
  W = W_{\rm MSSM} + \epsilon_{ij} { \hat{N}}_R {Y}_N { \hat{l}}^i_{L} {\hat H}^j_2\,.
  \label{eq:superpotential}
\ee
Here, $i$ and $j$ are SU(2) indices, and $\hat{l}_{L}$ are the left-handed lepton 
superfields. We include only terms conserving lepton flavor in each family; this is not crucial
at any step for our results, but simplifies the discussion. Under this assumption, the
$3\times3$ Yukawa matrix ${\bf Y}_N$ in Eq.~(\ref{eq:superpotential}) is diagonal; at the
same time, we can add only two extra-terms in the soft-SUSY breaking potential:
\be
  V_{{\rm soft}} =  V_{{\rm soft \; MSSM}} \;
  - \epsilon_{ij} {\bf \tilde{N}}^{*}_R {\bf A}_N {\bf Y}_N {\bf \tilde{l}}^i_{L}  H^j_2
 + {\bf \tilde{N}}_{R}^{*} {\bf M}_{N}^{2} {\bf \tilde{N}}_{R}\,,
  \label{eq:Vsoft}
\ee
with ${\bf A}_N$ and ${\bf M}_N$ being diagonal matrices. For each generation, a Dirac neutrino 
mass term
\be
  m_\nu \equiv Y_N \langle H_2^0 \rangle = Y_N \frac{\sqrt{2} M_W}{g} \sin\beta\
\ee
is induced, while sneutrinos, in the basis ($\tilde{\nu}_L$, $\tilde{N}_R$), acquire the mass matrix  
\be
  {\cal M}^2 =  \left( \begin{array}{cc}
  M_L^2 + m_\nu^2 + D_\nu &
  m_\nu ( A_N^* - \mu^* \cot\beta ) \\
  m_\nu (A_N - \mu \cot\beta ) &
  M_N^2 + m_\nu^2 \\
  \end{array} \right)\;.
  \label{eq:snumass}
\ee
Since the off-diagonal terms are proportional to $m_\nu$, the sneutrino mass eigenstates,
which we will denote by $\tilde{\nu}$ and $\tilde{N}$, essentially coincide with the interaction eigenstates; the
mixing angle $\theta$ is given by:
\be
  \tan (2\theta) = \frac{2\,(A_N - \mu \cot\beta )}{(M_L^2 + D_\nu - M_N^2)} m_\nu\;,
\ee
i.e.
\begin{eqnarray}
   \label{eq:theta}
   \cos\theta & \simeq & 1 \\ \nonumber
   \sin\theta & \simeq & \frac{\sqrt{2} M_W \,(A_N - \mu \cot\beta )}{(M_L^2 + D_\nu - M_N^2)} 
   \frac{Y_N}{g} \equiv R  \frac{Y_N}{g} \;.
\end{eqnarray}
The dimensionless ratio $R$ is introduced here as the relevant combination of the unknown
terms in the mass matrix in Eq.~(\ref{eq:snumass}); for SUSY parameters at the TeV scale, 
$R$ is expected to be of order $10^{-1}$. A much larger value has been considered in 
Ref.~\cite{Asaka:2005cn}, in the limit of degeneracy of the parameters $M_L$ and $M_N$ 
in the denominator of $R$: a large $R$  enhances the production of right-handed sneutrinos 
from the decay of SUSY particles when, in the very early Universe, the latter are relativistic 
and in thermal equilibrium. We are interested here in the opposite regime, i.e. the limit in which
the abundance of right-handed sneutrinos is negligible prior the decay of thermal relic 
neutralinos; hence, in the discussion below, the parameter $R$ is assumed of order $10^{-1}$
or smaller.

We suppose now that one of the 3 $\tilde{N}$ is the LSP and that the lightest neutralino 
is the NLSP. As in the case that we have studied in the previous Section, dark matter will then be generated in the decay of long-lived 
thermal-relic NLSPs into LSPs:
\be
\tilde{\chi}^0_1  \rightarrow  \tilde{N} + \bar{\nu}\,,
\ee
where $\tilde{\chi}^0_1$ denotes the lightest neutralino. The corresponding decay rate,
\be
  \Gamma=\frac{\left| g_{\tilde{\chi} \tilde{N} \nu} \right|^2}{32\pi} 
  \left[ 1 - \left(\frac{M_{\tilde{N}}}{M_{\tilde{\chi}}}\right)^2 \right]^2 \, M_{\tilde{\chi}},
\ee
takes the form of Eq.~(\ref{Gamma1}), with
\be
  \left|\tilde g_\mathrm{eff}\right|^2 = \frac{(1+\delta/2)^2}{(1+\delta)^4} 
  \left| g_{\tilde{\chi} \tilde{N} \nu} \right|^2 \,.
  \label{geffsnu}
\ee
For the neutralino-sneutrino-neutrino coupling that appears in the above expressions, we find
\be
   g_{\tilde{\chi} \tilde{N} \nu} =    
   \left( g^\prime N_{11} - g N_{12} \right) \sin\theta - Y_N N_{14} \cos\theta\,,
\ee
where we have implemented the standard projection of the lightest neutralino on the 
interaction basis:
\be
  \tilde{\chi}^0_1= 
  N_{11} \tilde{B} + N_{12} \tilde{W}^3 + 
  N_{13} \tilde{H}^0_1 + N_{14} \tilde{H}^0_2\,.
\ee
 As opposed to the gravitino dark matter case,  
there is thus an explicit dependence of the decay rate not only on $\delta$ and $M_{\tilde{\chi}}$ but also on the
SUSY parameters setting neutralino mass and mixing, as well as those entering in the 
dimension-less ratio $R$ (In particular, one should notice that all contributions to 
$g_{\tilde{\chi} \tilde{N} \nu}$ are linear in the Yukawa coupling $Y_N$).

Morover, in the computation of the neutralino relic densities, results are in general sensitive to a larger 
number of MSSM parameters than in the case we discussed before for a left-handed sneutrino. 
We are mainly interested in understanding, for a given neutralino mass, what kind of neutralino
composition is compatible with the relic abundance constraint. 
In this respect, it is sufficient to consider a simplified scheme, covering the full range of neutralino compositions: the split SUSY framework~\cite{Arkani-Hamed:2004fb,Giudice:2004tc} in which
all scalar superpartners are assumed to be much heavier than Gauginos and Higgsinos.
As discussed in Ref.~\cite{Masiero:2004ft} the DM phenomenology of the model can be described 
in terms of the Bino, Wino and Higgsino mass parameters, respectively, $M_1$, $M_2$ and $\mu$. 
Since sfermions are heavy, neutralino pair-annihilation rates are dominated by gauge boson final 
states; these are at full strength for Higgsinos and Winos, while are strongly suppressed for pure 
Binos. To reproduce the measured dark matter density one has to modulate the mixing between 
these states in the lightest neutralino and/or adjust coannihilation effects with other fermionic
superpartners. If $M_2$ is the lightest SUSY parameter, we recover the case of the pure Wino, 
which as mentioned above has a thermal relic density matching the observed value for a mass 
of about 2.2~TeV. On the other hand, if $\mu$ is the lightest parameter, and consequently the
lightest neutralino is a pure Higgsino, the annihilation rate is slightly smaller and the cosmological
bound is saturated at  about 1.1~TeV. Models with a Higgsino-Wino mixing cover the mass range
between the pure states. Introducing a Bino component in the lightest neutralino allows
to find configurations with lighter masses (essentially as light as the W-boson); again there 
are two possibilities: if the $\mu$ parameter is of the order of $M_1$, there is a large Bino-Higgsino
mixing modulating the annihilation rate and hence the relic density. On the other hand
if $\mu$ is heavy and $M_2$ light, since the transition between Bino and Wino LSP is very 
sharp~\cite{Masiero:2004ft}, one needs to consider a configuration with $M_2$ just slightly
heavier than $M_1$: the LSP is a very pure Bino and coannihilations with Winos play the key
role in thermal decoupling (see~\cite{Masiero:2004ft} for further details). Had we introduced
light sfermions in our framework ,  Binos could have efficiently annihilated into heavy 
quarks or leptons, and we would have inferred a different Bino component in the LSP;
nevertheless, the LSP composition would still have been within the range of extreme 
Gaugino-Higgsino fractions we find in split SUSY.

\begin{figure}[t!]
   \includegraphics[width=\columnwidth]{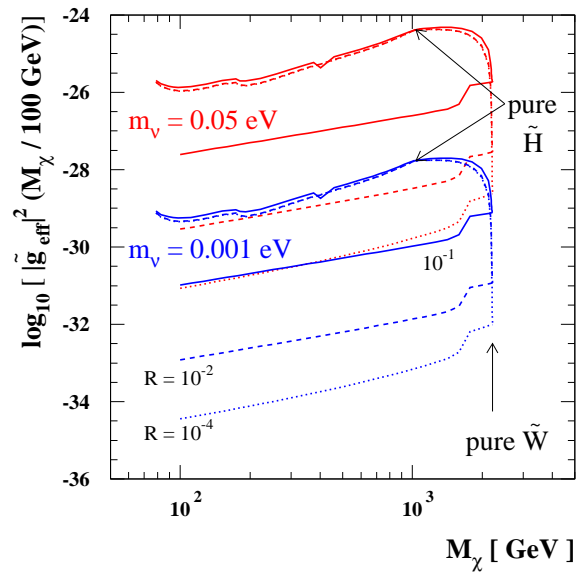}
     \caption{Coupling strength versus neutralino mass for models with DM density matching the
    DM observed value. The limit of small mass-splittings between the lightest neutralino 
    and the right-handed sneutrino is assumed. Three sample values of the parameter
    $R$ ($R=10^{-1},\,10^{-2},\,10^{-4}$ for, respectively, the solid, dashed and dotted curves) 
    and two for the  Dirac neutrino mass term $m_\nu$ (0.05~eV and 0.001~eV, respectively,
    for upper and lower curves) are considered. Compare with Fig.~\protect{\ref{fig_particle}}
    and note that the values on the vertical axis which are relevant to address the small scale
    problems are of the order of  $10^{-31}$ or smaller. 
}
     \label{fig:geff2}
\end{figure}

In Fig.~\ref{fig:geff2} we consider the limit of small mass-splittings between the lightest neutralino 
and the right-handed sneutrino, and select models with neutralino thermal relic abundance 
matching the measured DM density; for such models, we plot the effective coupling squared introduced 
in Eq.~(\ref{geffsnu}) times the neutralino mass (i.e. the quantity we introduced in Fig.~\ref{fig_particle}), 
versus the neutralino mass itself. We show results for three sample values of the dimensionless
ratio $R$ (the solid, dashed and dotted curves in Fig.~\ref{fig:geff2} correspond, respectively, 
to $R=10^{-1},\,10^{-2},\,10^{-4}$), and for each of these we consider the three regimes 
mentioned above: the upper branches of each curve (which nearly overlap for the three values 
of $R$) start at small neutralino mass in the regime of  Bino-Higgsino mixing, reach the pure
Higgsino configuration and continue down to the pure Wino case; the lower branches starts at the 
heavy mass end with pure Winos, make the transition into pure Binos (the step along each 
curve)  and a progressive tuning in Bino-Wino coannihilation effects allow them to extend down 
to light neutralino masses. 

Two sample values for the Yukawa coupling $Y_N$, corresponding, respectively, to a Dirac 
neutrino mass term $m_\nu$ of  0.05~eV and 0.001~eV, are displayed in Fig.~\ref{fig:geff2}
for illustrative purposes. Since, as we already mentioned, all contributions to the coupling 
$g_{\tilde{\chi} \tilde{N} \nu}$ are linear in the Yukawa coupling $Y_N$, the results in the 
plot just scale with the square of $m_\nu$. 
The overall neutrino mass scale is not known. The upper  bound from Cosmology is at the level of
0.3~eV~\cite{Seljak}. On the other hand, the mass-squared differences among neutrinos have been 
determined with good accuracy in neutrino-oscillation experiments, the largest being for neutrinos 
taking part in atmospheric oscillations~\cite{Fogli:2005cq}:
\begin{equation}
\Delta m_\nu^2 \simeq \left[0.05 \; {\rm eV}\right]^2 \,.
\end{equation}
This sets a lower bound on the mass of the heaviest neutrino. In the scenario in which
lepton-number is strictly conserved and there is no neutrino Majorana mass terms, the 
largest $Y_N$ (which is the relevant one for our discussion, since it induced the fastest decay 
mode) has then to induce a Dirac mass term $m_\nu$ at least at the level of $0.05 \; {\rm eV}$.
In such case, the values we obtain in Fig.~\ref{fig:geff2} for the effective coupling 
squared times neutralino mass are well above those we found in  Fig.~\ref{fig_particle} 
are needed for a solution to the small-scale problems, namely 
$ \left|\tilde g_\mathrm{eff}\right|^2 M_\chi/100$~GeV not exceeding  $10^{-31}$. 
This scenario gives viable DM  candidates, however these are not relevant (or at most marginally 
relevant) for addressing the problems in CDM structure formation on small scales. 

On the other hand, if we consider a slightly smaller Dirac neutrino mass term, i.e. if we assume 
there are Majorana mass terms contributing to the pattern of neutrino masses as measured in 
neutrino oscillation experiments, our predictions can cover the entire region of the parameter 
space which is relevant to solve both small-scale structure problems at the same time,
provided the mass splitting between the long-lived neutralino and the DM right-handed
sneutrino is at the percent level or smaller. 
Models with two heavy right-handed sneutrino and one light $\tilde{N}$ with a Yukawa coupling 
$Y_N$ as small as the one needed in our contest, can indeed arise naturally in SUSY 
frameworks~\cite{francesca}. On the other hand, it is harder to explain the degeneracy in mass
between quasi-stable and stable species, and we have unfortunately to rely on a certain amount 
of tuning  of the parameters in the model. In the limit of larger mass splittings we could argue 
again, as we did for gravitino DM that the model can address and solve the problem of 
the overabundance of satellites in Milky Way size galaxies; at the same time, however, the 
problem of limiting the amount of radiation or hadronic components below about 0.5\% of
the energy released in the decay reappears as well.

\subsection{The analogous cases in universal extra dimensions}
\label{UED}

Models with universal extra dimensions (UED) \cite{Appelquist:2000nn}, where all standard model (SM) fields are allowed to propagate in a higher-dimensional bulk, have received a great deal of attention since it was realized that they naturally give rise to a new class of dark matter candidates \cite{Cheng:2002iz,Servant:2002aq}: the higher-dimensional extra degrees of freedom appear in the low-energy effective 4D theory as towers of new, heavy states, the lightest of which -- similar to the case of $R$-parity in supersymmetry -- is stable due to an internal $Z_2$ symmetry (this ``KK-parity'' appears as a remnant of the higher-dimensional translational invariance after the orbifold compactification); thermally produced in the early universe, the lightest Kaluza-Klein particle (LKP) acquires  the right WMAP relic density for a compactification scale of about $R^{-1}\sim1\,$TeV \cite{Kong:2005hn,Burnell:2005hm,Kakizaki:2006dz}. In fact, the analogy to supersymmetric models goes even further for energies close to this scale, when only the lightest state of each KK-tower is kinematically accessible. In that situation, every SM particle effectively comes equipped with only \emph{one} massive partner, just as in the supersymmetric case (having, however, the \emph{same} spin). A clear discrimination  between these two models at colliders may therefore actually be a rather challenging task; for this reason, the UED model has sometimes also been dubbed ``bosonic supersymmetry'' \cite{Cheng:2002ab,Datta:2005zs,Battaglia:2005zf}.

With these introductory remarks in mind, it should not come as a surprise that SuperWIMPs also appear in the UED setup. What is more, we can expect them to be the exact analogues of the supersymmetric cases that we discussed above -- i.e. the Kaluza-Klein graviton, $G^{(1)}$, and the first Kaluza-Klein state of the right-handed neutrino, $\nu^{(1)}_R$, respectively. Before we continue to discuss these SuperWIMP DM candidates in turn, let us stress a particular feature about extra-dimensional models, namely that one is generically driven to small mass splittings $\delta$. This is because the masses of all KK states are degenerate at tree level; taking into account radiative corrections, they take the form
\be
  M_{n}^2=\left(\frac{n}{R}\right)^2+m_\mathrm{SM}^2+\delta M_n^2\,,
\ee
where $n$ is the KK number of the state, $m_\mathrm{SM}$ the corresponding electroweak (SM) mass and $\delta M$ the mass shift due to radiative corrections (here, and in the following, we make the simplifying assumption of only one extra dimension, compactified on $S_1/Z_2$). For KK partners to SM particles, a naive estimate for the radiative corrections would be
\be
 \label{deltanaive}
 \delta M_n^2\sim\alpha_i M_{n}^2\,,
\ee
where $\alpha_i$ is the relevant gauge coupling constant; one therefore expects $\delta M\gg m_\mathrm{SM}$ for TeV compactification scales, at least for fermions (with the possible exception of the top quark). For the $G^{(1)}$ or the $\nu^{(1)}_R$, on the other hand, one can neglect radiative corrections to a very good approximation; in this case, the tree-level degeneracy is not lifted and we have $M_{G^{(1)}}\simeq M_{\nu^{(1)}_R}\simeq1/R$. Taking these considerations at face value, one would thus, in most cases, naturally find $\delta\sim\alpha_i/2\lesssim0.01$ for the \emph{smallest} WIMP - SuperWIMP mass splittings -- which, from our discussion in Section \ref{sec_setup}, is a crucial ingredience for a possible solution to the small-scale problems of standard CDM cosmology.

\begin{figure}[t]
       \includegraphics[width=\columnwidth]{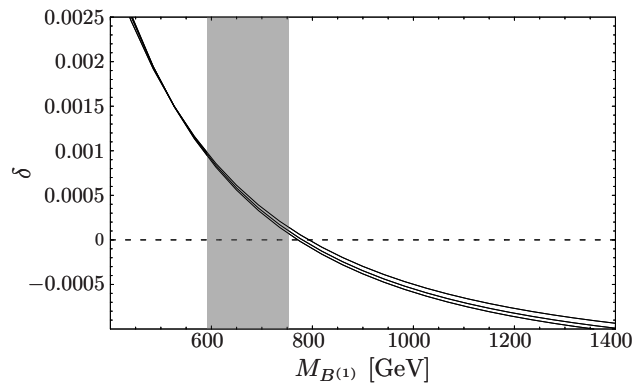}
     \caption{This figure shows the relative difference between the $B^{(1)}$ mass and the inverse compactification scale, $\delta\equiv R\,M_{B^{(1)}} -1$, as computed in the mUED setup \cite{Cheng:2002iz}. The grey band shows the region that is consistent with the $2\sigma$ WMAP relic density constraint for a Higgs mass $m_h\lesssim150\,$GeV; for higher Higgs masses, the grey region broadens and shifts to the right, allowing inverse compactification radii up to $R^{-1}\lesssim1.3\,$TeV \cite{Kakizaki:2006dz}. The different curves correspond to cutoff scales $\Lambda R=20,30,40$.
}
     \label{fig_B1}
\end{figure}

The actual situation, however, is complicated by two further effects. First, radiative corrections can be both positive and negative and, second, they receive cutoff-dependent contributions from counterterms localized at the orbifold fixpoints. For simplicity, one usually adopts the self-consistent assumption that these boundary terms are small at the cutoff scale $\Lambda$ \cite{Cheng:2002iz}. In this approach, which is often referred to as the \emph{minimal} UED model (mUED), the lightest SM partner is the $\gamma^{(1)}$.
Since the ``Weinberg angle''  (i.e. the rotation angle) for KK modes is essentially driven to zero, the LKP is well approximated by the $B^{(1)}$, the first KK excitation of the weak hypercharge gauge boson. As shown in Fig.~\ref{fig_B1}, it receives slightly positive mass corrections, $10^{-4}\lesssim\delta\lesssim10^{-3}$, for compactification scales of cosmological relevance. 

The $B^{(1)}$ in the mUED model is therefore actually not the lightest, but only the \emph{next} to lightest KK particle (NLKP), the lightest being the KK graviton $G^{(1)}$.\footnote{
For high values of the standard model Higgs mass, $m_h\gtrsim170\,$GeV, the $B^{(1)}$ can be the actual LKP and (for $R\gtrsim800\,$GeV) constitutes a viable DM candidate that evades the over-production of photons discussed below \cite{Cembranos:2006gt}. For even higher values,  $m_h\gtrsim250\,$GeV,
one would actually have a charged  LKP (or NLKP) \cite{Cembranos:2006gt}; for relic densities even remotely close to the required amount of DM, however, this possibility is cosmologically excluded and we will therefore not consider it in the following.}
 The latter thus appears as a typical SuperWIMP DM candidate that may arise from the late decay of the $B^{(1)}$ \cite{Feng:2003nr,Shah:2006gs}  (Note that also the thermal production of the $G^{(1)}$ is much more efficient than that of the $\tilde G$, which puts severe constraints on the reheating temperature).  The only allowed (two-body) decay mode is $B^{(1)}\rightarrow G^{(1)}\gamma$, which results in a $B^{(1)}$ lifetime of \cite{Feng:2003nr}
\be
 \label{B1Gg}
 \tau\approx2.9\times10^{13}\,\delta^{-3}(M_{B^{(1)}}/\mathrm{GeV})^{-3}\,\mathrm{s}\,,
\ee
where $\delta$ and $M_{B^{(1)}}$ are subject to the relic density constraint shown in Fig.~\ref{fig_B1}. However, as can easily be seen from Fig.~\ref{fig_constraints}, the photons from the decay would  lead to a distortion of the CMB that is clearly inconsistent with observations (this conclusion may be evaded for extremely tiny $\delta$, associated to $\tau\sim t_0$ \cite{Cembranos:2007fj}).

A possible way out is to leave the somewhat arbitrary framework of the mUED model and take the KK masses as  free parameters that may be varied around their respective mUED values (see, e.g., \cite{Servant:2002aq,Kong:2005hn,Burnell:2005hm}). One can then consider a setup in which the lightest SM partner is the $\nu_L^{(1)}$, the KK excitation of the neutrino \cite{Servant:2002aq}. Such a particle is ruled out as a DM candidate by  direct detection experiments \cite{Servant:2002hb}, but these bounds of course do not apply if it has already decayed into the $G^{(1)}$. This time, the only allowed two-body decay mode is $\nu_L^{(1)}\rightarrow G^{(1)}\nu$, in accordance with the general requirements derived in Section \ref{sec_setup}. The decay rate is then given by~\cite{Feng:2003nr}:
\be
\Gamma_{\nu_L^{(1)}}=\frac{1}{48M^2_\mathrm{Pl}}\frac{M^7_{\nu_L^{(1)}}}{M^4_{G^{(1)}}}\left(1-\frac{M^2_{G^{(1)}}}{M^2_{\nu_L^{(1)}}}\right)^4\left(2+3\frac{M^2_{G^{(1)}}}{M^2_{\nu_L^{(1)}}}\right)\,,
\ee
so it takes the form anticipated in (\ref{Gamma2}), with
\be
  \left|\tilde g_\mathrm{eff}\right|^2=5\, \frac{(1+\delta/2)^4}{(1+\delta)^4}\left(1+\frac{4}{5}\delta+\frac{2}{5}\delta^2\right)\,.
\ee
Correcting for a factor of $\left|\tilde g_\mathrm{eff}\right|^{2/3}\approx1.7$ for small mass splittings, we can now use Fig.~\ref{fig_particle_b} to directly read off the KK neutrino mass that is required in order to reach the dark shaded region, where both small scale problems can be resolved. In particular, we find that a mass of at least $M_{\nu_L^{(1)}}\approx2.5\,$TeV is needed in order to reach the parameter region of interest (this corresponds to the largest possible mass splitting, $\delta\sim0.02$, for solving the small scale problems); going to smaller mass splittings, we see that for, e.g., $\delta=0.005$ (where the necessary fine-tuning in the mass starts to become less severe) one even needs $4.6\,\mathrm{TeV}\lesssim M_{\nu_L^{(1)}}\lesssim6.5\,\mathrm{TeV}$. Since a $\nu_L^{(1)}$ (N)LKP is expected to have roughly the same relic density as a $B^{(1)}$ (N)LKP \cite{Servant:2002aq}, it seems extremely unlikely that one can drive masses to such high values, even when including very efficient coannihilation channels by a suitable tuning of KK masses \cite{Kong:2005hn}. The KK graviton is thus excluded as a solution to \emph{both} small scale problems of standard CDM. Note that, due to the strict bounds on the allowed size of free-streaming effects, the KK graviton from the late decay of KK neutrinos is actually ruled out as a DM candidate for most of the parameter space -- unless one allows for rather large masses and mass splittings. For $\delta\sim0.1$,~e.g., the Lyman-$\alpha$ bound can be evaded for masses $M_{\nu_L^{(1)}}\gtrsim1.0\,$TeV.

A second possibility to evade the KK graviton problem of the mUED model is to introduce a SM Dirac neutrino \cite{Matsumoto:2006bf} (alternatively, one may also consider a situation in which gravity propagates in more dimensions than the SM fields; in this case, one can adjust the KK graviton to becomes more massive than the SM KK modes \cite{Dienes:2001wu}, so that the $B^{(1)}$ would be the LKP and thus stable). The right-handed neutrino then receives a tower of heavy KK modes in the same way as its left-handed counterpart. Though tiny, the non-zero mass $m_\nu$ of the Dirac neutrino leads to a mixing of the right- and left-handed neutrino KK states. This induces the following effective coupling between the $B^{(1)}$ and the first KK mode of the right-handed neutrino, $\nu_R^{(1)}$:
\be
 i\frac{g_Y}{2}\sin\alpha \,\gamma^\mu P_L\,,
\ee
where $P_L=(1-\gamma^5)/2$ is the usual left-handed projection operator and the mixing angle $\alpha$ is given by
\be
  \tan2\alpha=\frac{2m_\nu}{M_{\nu_L^{(1)}}+M_{\nu_R^{(1)}}}\approx2\alpha\,.
\ee
Due to this coupling, a new decay channel
\be
 B^{(1)}\rightarrow\nu_R^{(1)}\bar\nu+\bar\nu_R^{(1)}\nu
\ee
 opens up, with a decay rate
\be
   \label{B1Nn}
  \Gamma_{B^{(1)}}=\frac{g_Y^2}{96\pi} \sin^2\alpha\, M_{B^{(1)}}\left(1-\frac{M_{\nu_L^{(1)}}^2}{M_{B^{(1)}}^2}\right)^2\left(2+\frac{M_{\nu_L^{(1)}}^2}{M_{B^{(1)}}^2}\right)\,.
\ee
This is of the form (\ref{Gamma1}), with
\be
  \label{gnuR}
  \left|g_\mathrm{eff}\right|^2=\frac{g_Y^2}{8\pi}\left(\frac{m_\nu}{M_{\nu_L^{(1)}}+M_{\nu_R^{(1)}}}\right)^2\,\left[1-\frac{8}{3}\delta+\mathcal{O}(\delta^2)\right]\,.
\ee
Since the $\nu_R^{(1)}$ does not receive any appreciable radiative corrections, its mass is to a very good approximation given by $M_{\nu_R^{(1)}}=R^{-1}$ and the quantity $\delta$ is the same as that shown in Fig.~\ref{fig_B1}. Comparing now Eq.~(\ref{B1Gg}) with Eq.~(\ref{B1Nn}), we observe that
\be
 \frac{\Gamma_{B^{(1)}\rightarrow\nu_R^{(1)}\bar\nu}}{\Gamma_{B^{(1)}\rightarrow G^{(1)}\gamma}}= 5.1\,\delta^{-1}\left(\frac{M_{B^{(1)}}}{\mathrm{TeV}}\right)^{-4}\left(\frac{m_\nu}{0.01\,\mathrm{eV}}\right)^2\,.
\ee
Since atmospheric neutrino experiments place a lower limit of $m_\nu\gtrsim0.05\,$eV on the heaviest neutrino \cite{Fogli:2005cq}, the KK graviton problem is thus easily solved in this setup.

\begin{figure}[t]
       \includegraphics[width=\columnwidth]{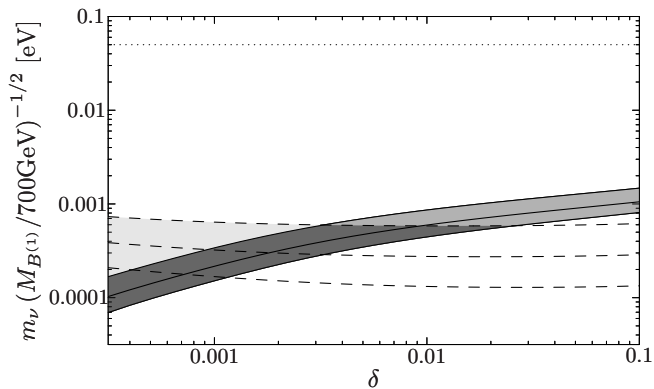}
     \caption{The shaded regions show the range of parameters where a setup with DM from the late decay $B^{(1)}\rightarrow\nu_R^{(1)}\bar\nu$ can solve (at least one of) the CDM small scale problems; the area below the shaded regions is excluded by the data from the Lyman-$\alpha$ forest. The dotted line represents the experimental \emph{lower} bound on the neutrino mass, $m_\nu\gtrsim0.05\,$eV, for a compactification scale of $R^{-1}=700\,$GeV.
}
     \label{fig_nuR}
\end{figure}

Let us now investigate whether a right-handed KK neutrino may also solve the CDM small scale problems. To this end, we use Eq.~(\ref{gnuR}) and show in Fig.~\ref{fig_nuR} the parameter plane $\delta$ - $m_\nu$ as a rescaled version of Fig.~\ref{fig_particle}. As becomes obvious from this figure, the $\nu_R^{(1)}$ as a SuperWIMP DM candidate has virtually no impact on the small scale problems -- simply because the decay rate is too large. Even if one leaves the mUED model, where the $B^{(1)}$ would become the LKP for compactification scales $R^{-1}\gtrsim800\,$GeV, unrealistically high (N)LKP masses are needed to sufficiently suppress the couplings so as to approach the parameter region of interest (Note that we have an effective scaling with masses as $\Gamma\propto m_\nu^2/M_{B^{(1)}}$). Unless one finds a way to construct a
 scheme where the main contribution of the observed neutrino mass pattern does not derive from the Dirac mass terms that we have introduced here, a solution of the CDM small-scale problems with a $\nu_R^{(1)}$ DM candidate is thus not possible within our general framework -- although, contrary to the supersymmetric case, we naturally expect the required small mass splittings for such a setup.

\section{Summary and conclusions}
\label{sec_conc}

We have re-considered the possibility that the main part of the present dark
matter component in the Universe derives from the late decay of a relic
population of (cold) quasi-stable particles. While not changing standard CDM
cosmology on large scales, such a scenario introduces interesting new
effects on small scales; it has therefore been advocated as a possible
solution to the problems that current N-body simulations in $\Lambda$CDM cosmologies are facing,
viz. the overabundance of halo substructures on the one hand, and inner halo profiles that
are too concentrated and steeper than what is suggested by observations on the other hand. Since its first
proposal, this idea has attracted considerable attention, not the least  as
well-motivated DM scenarios of this type seemed to arise
naturally in supersymmetric or extra-dimensional extensions to the standard
model.

In the first part of this paper, the general requirements for solving the
two small scale problems in such a way have been presented in a form that
makes it straightforward to check whether any given particle physics model
fits into this scheme, taking into account the various relevant
astrophysical constraints. In the second part, we have then applied our
general discussion to those DM candidates that have often been quoted in
this context as ``naturally'' satisfying the necessary requirements; the
supersymmetric gravitino and the Kaluza-Klein graviton in
theories with universal extra dimensions. We find that  these DM candidates 
are actually \emph{not} suited to solve both small scale problems \emph{simultaneously}.
This contradiction with previous claims is mainly related to the fact that, in all 
our explicit models, we take as requirement for naturalness the hypothesis
that DM is generated in the decay of thermal relics from the early Universe,
as opposed to other arbitrary and ad-hoc initial conditions.

As an alternative, we have introduced here the scenario with DM in the form of 
a right-handed sneutrino or of the first Kaluza-Klein state of a right-handed neutrino.
These play again the role of the superWIMP, since their interactions are mediated
by a very small Yukawa coupling; moreover, the mechanism we propose is still DM
production in the Universe through the decay of quasi-stable thermal relics. 
We have found that, in region of the parameter space relevant to solve the small-scale 
structure problems, the induced Dirac neutrino mass terms are slightly smaller than the
minimum neutrino mass scale required by neutrino oscillation experiments. 
A viable scenario, including Majorana neutrino mass terms and one subdominant Dirac 
mass term, can be naturally embedded in a supersymmetric framework; on the other
hand, some fine-tuning in the parameter space seems unavoidable to reproduce 
the required small (i.e. at the percent level or smaller) mass splitting between the long-lived 
species, namely the lightest neutralino, and the DM right-handed sneutrino. 
For a right-handed Kaluza-Klein neutrino in theories with universal extra dimensions, 
a small mass splitting between WIMP and SuperWIMP is expected, still we miss a 
detailed picture for the generation of neutrino masses.

In conclusion, we have shown that a solution to the small scale structure problems can be 
achieved in the framework in which DM is generated in the decay at late times of a 
quasi stable CDM particles. At the same time, this solution is pointing to very specific 
features in the underlying particle physics model, in particular concerning mass splittings
and coupling strengths, and a certain amount of fine-tuning seems intrinsic in scenarios of this 
kind.

\begin{acknowledgments}

P.U. would like to thank Julien Lesgourgues for interesting discussions,
and the Galileo Galilei Institute for Theoretical Physics for the hospitality 
during a period in which part of this work was developed. 
The work of T.B.\ and P.U.\ was supported by the Italian INFN under
the project ``Fisica Astroparticellare'' and the MIUR PRIN ``Fisica
Astroparticellare''.  P.U. was also partially supported by the EU 6th Framework 
Marie Curie Research and Training network "UniverseNet" (MRTN-CT-2006-035863).

\end{acknowledgments}

\end{document}